\documentclass[aps,twocolumn]{revtex4}

\input epsf
\def\plotone#1{\centering \leavevmode
\epsfxsize= 1.0\columnwidth \epsfbox{#1}}

\def\be{\begin{equation}}
\def\ee{\end{equation}}
\def\bea{\begin{eqnarray}}
\def\eea{\end{eqnarray}}

\def\cmm2{{\,\rm cm^{-2}}}
\def\cm2{{\,{\rm cm}^2}}
\def\cmm3{{\,{\rm cm}^{-3}}}
\def\gcmm3{{\,{\rm g\,cm^{-3}}}}

\def\fun#1#2{\lower3.6pt\vbox{\baselineskip0pt\lineskip.9pt
  \ialign{$\mathsurround=0pt#1\hfil##\hfil$\crcr#2\crcr\sim\crcr}}}
\def\C{{\cal C}}

\def\p3m{P$^3$M}

\def\ga{\mathrel{\mathpalette\fun >}}
\def\fun#1#2{\lower3.6pt\vbox{\baselineskip0pt\lineskip.9pt
  \ialign{$\mathsurround=0pt#1\hfil##\hfil$\crcr#2\crcr\sim\crcr}}}

\def\cldd{\C_l^{dd}}

\newcommand{\tableskip}{\\[-6pt]}

\begin{document}
\bibliographystyle{prsty}
\title{Future Probes of the Primordial Scalar and Tensor Perturbation Spectra:

Prospects from the CMB, Cosmic Shear and High-Volume Redshift Surveys}
\author{Lloyd\ Knox}
\affiliation{Department of Physics, One Shields Avenue\\
University of California, Davis, California 95616, USA}
\date{\today}

\begin{abstract}
Detailed study of the scalar and tensor perturbation spectra can
provide much information about the primordial fluctuation-generator,
be it inflation or something else.  The tensor perturbation spectrum
may be observable through its influence on CMB polarization, but only
if the tensor-to-scalar ratio, $r \equiv T/S$, is greater than about
$10^{-5}$.  The tensor tilt can be measured with an error of
$\sigma(n_T)$ that decreases with $r$ from $0.1$ at $r=0.001$ to
$0.02$ at $r = 0.1$. Current CMB constraints on the
scalar perturbation spectrum can be improved by higher--resolution CMB
observations and/or by tomographic cosmic shear observations.  These
can both shrink errors on the tilt ($n_S$) and running ($n_S' \equiv
dn_S/d\ln k$) to the $10^{-3}$ level.  Stunning as these results would
be, it may become very desirable to improve upon them an order of
magnitude further in order to study the expected departures from $n_S'
= 0$.  Such improvements are likely to require observation of
three--dimensional clustering over very large volumes.  Unfortunately,
to get down to the $10^{-4}$ level will require a sparse spectroscopic
redshift survey with about $10^9$ galaxies spread over a volume less
than but comparable to that of the observable Universe.
\end{abstract}
  \maketitle

\section{Introduction}

Inflation is doing remarkably well with respect to observations.  The
evidence that structure in the Universe formed from adiabatic, nearly
scale-invariant fluctuations is very strong \cite{bennett03}.  The mean
spatial curvature has been determined with high accuracy and is
consistent with zero \citep{spergel03}.  The perturbations are highly
Gaussian \cite{komatsu03}.  These are all predictions of inflation, of
varying degrees of robustness.  Topological defect models, formerly
inflation's chief competition, have been strongly ruled out 
(e.g., \cite{knox00}).

Despite this great success we still have little understanding of the
physics that led to these initial perturbation spectra.  There are
many different models of inflation.  As Michael Turner said in his
talk at this meeting, we have a Landau-Ginzburg theory, but we are
missing the underlying BCS theory.

To discriminate among inflationary models, to assist in the quest for
a BCS theory of inflation, and perhaps to discriminate between
inflation and alternatives (such as we heard from Steinhardt and from
Durrer), we want to take the determination of the primordial tensor
and scalar perturbation spectra to a qualitatively new level of
precision.  This point was also emphasized in Albrecht's talk.

I will first briefly review the current constraints on the scalar
perturbation spectra before turning to the future.  The discussion of
future experiments will start with the constraints on the tensor
perturbation spectra from CMB experiments.  We will then look
at how current CMB data can be complimented either with higher--resolution
CMB data and/or tomographic cosmic shear.  Finally, I consider results
that can in principle be achieved by very high-volume spectroscopic redshift 
surveys.

\section{Non-zero running?}

A question of great relevance to inflationary models is the
significance of evidence for running of the spectral index.  The
evidence from CMB data alone is very weak: $n_S' \equiv dn_s/d\ln k =
-0.055 \pm 0.038$ \cite{spergel03}.  Combining CMB data with the Croft
et al. matter power spectrum inferred from high--resolution
observations of the Ly$_\alpha$ forest results in $n_S' = -0.031 \pm
0.017$ \cite{spergel03}.  Other authors \cite{seljak03} working with
the same datasets have since found much looser constraints on $n_S'$.
The looser bound on $n_S'$ is due to their marginalization over the
mean ionizing flux as a function of redshift.  The importance of
marginalizing over this parameter, which leads to a large degeneracy
between spectral index and amplitude, was pointed out in
\cite{zaldarriaga01}.

We had much discussion at this meeting about constraints from
combining CMB data with low--resolution spectra from the Sloan Digital
Sky Survey (SDSS), as presented by Hui and by Seljak.  
Although lower resolution, there are thousands of
these spectra allowing for very small statistical errors.  Possible
systematic errors have not yet been understood and controlled well
enough to allow for any firm detections of $n_S' \ne 0$.

To summmarize the current situation: there is no strong evidence for
$n_S' \ne 0$, but very interesting results may be coming soon from
SDSS quasar spectra combined with WMAP.

\section{CMB and Tensor Perturbations}

One can decompose a polarization pattern on the sky into curl--free modes
(E modes) and divergence--free modes (B modes) 
\cite{kamionkowski97,seljak97}.  Each scalar three-dimensional Fourier mode
only has one direction, that given by the wavevector ${\bf k}$, and therefore
only leads to E modes.  Tensor Fourier modes (gravitational waves) have
the direction given by ${\bf k}$ and also an orientation given by the
polarization of the gravitational wave.  They can therefore generate B modes
in addition to E modes.  Since, at least in linear perturbation theory, scalar
perturbations do not generate B modes, whereas tensor perturbations do, the
B mode has been proposed as a means to detect the tensor perturbations.

At second order in perturbation theory we cannot solve for just one
Fourier mode at a time and then sum up the resulting solutions; the
evolution of one Fourier mode is affected by the presence of others.
This mode-mode coupling leads to the generation of B modes even from scalar
perturbations \cite{zaldarriaga98}.  The dominant second order effect is 
gravitational lensing.  The lensing--induced B modes can obscure the 
gravitational wave contribution to the B modes.  If the lensing--induced 
B modes are not cleaned from the B-mode map, one can only detect the 
tensor signal (at $3 \sigma$) if $r > r_{\rm lim} = 1.7\times 10^{-4},
7\times 10^{-5}$ or, $2.4 \times 10^{-5}$ for $\tau = 0.05, 0.1$ or 0.2
respectively.  

Gravitational lensing leads to off-diagonal correlations in Fourier
space \cite{bernardeau97,zaldarriaga00a,okamoto02,cooray03}.  
These can be used to reconstruct the lensing potential \cite{hu01c,hu02b}
.  With
the lensing potential thus reconstructed, the maps can be unlensed.  A
perfectly unlensed map would have no scalar B modes in it.
Unavoidable imperfections in the lensing potential reconstruction mean
the unlensed map will have some residual B mode, even in the absence
of tensor perturbations.  These residual B modes prevent detection of
gravity waves unless $r > r_{\rm lim} = 1.4 \times 10^{-5}, 6 \times 10^{-6}$
or $2 \times 10^{-6}$ for $\tau = 0.05, 0.1$ or 0.2.  These 
limit calculations were first done by \cite{knox02} and \cite{kesden02}.

As shown in\footnote{Y.-S. Song \& L. Knox, in preparation}, 
if $r \ga 0.01$ then we can learn something about the shape of the
tensor B mode power spectrum.  The possible
constraints on $n_T$ as a function of $r$ are shown in Fig. 1.

\begin{figure}[htbp]
\label{fig:nT}
  \begin{center}
    \plotone{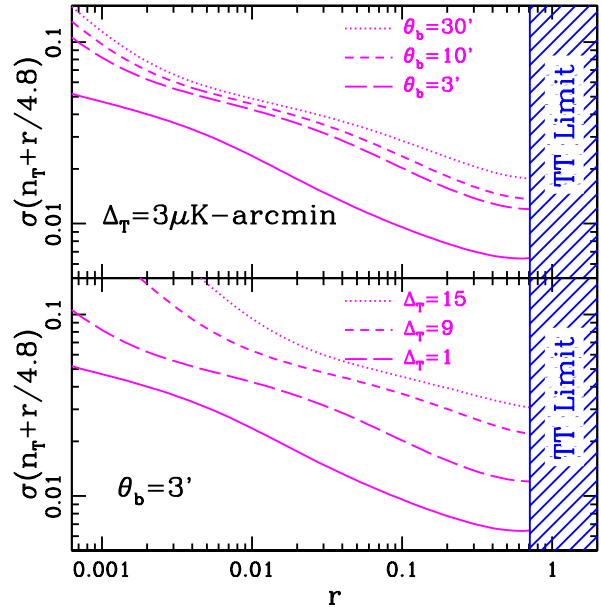}
    \caption{Error on $n_T+r/4.8$ from a full-sky, no-noise experiment
(solid line) and for varying angular resolutions (top panel) and varying
noise levels (bottom panel), as a function
of the tensor-to-scalar ratio, $r$.  The inflationary consistency
equation is $n_T + r/4.8=0$.  Values of $r >0.71$ are ruled out
by temperature power spectrum data. From Song \& Knox, in preparation.
}
\end{center}
\end{figure}

Tensor spectra from inflationary models with a single field in slow
roll obey the `consistency equation', $r=-5 n_T$ \cite{lidsey97}.
Note that, more generally, $r = f(\Omega_\Lambda) n_T$ and that $f(0)
\simeq -7$ and $f(0.7) \simeq -5$ \cite{knox95,turner95}.  From the
figure we can see that for $r \ga 0.03$ we can make a significant test
of the consistency relation.  The consistency equation applies for all
single-field models to first order in the slow roll parameters. 

\section{CMB and Scalar Perturbations}

Constraints to the scalar perturbation spectra can be improved by
pushing to higher angular resolution than the $\sim 13$' of the WMAP's
highest frequency channel.  It can not be improved arbitraily though
because of the exponential suppression of power that sets in at the
Silk damping scale, due to photon diffusion during recombination.
High sensitivity can to some degree fight against this exponential
cutoff, but not to indefinitely high $\ell$.

Here we show results expected for three different experiments (with parameters
specified in Table I for the
amplitude of the primordial gravitational potential power spectrum,
$P_\Phi^i(k_f)$, $n_S$, and $n_S'$ where \be \ln{[P_\Phi^i(k)]} =
\ln{[P_\Phi^i(k_f)]}+ \left(n_S(k_f)-3\right)\ln{(k/k_f)} +
n_S' [\ln{(k/k_f)}]^2.  \ee We also include the expected constraints on $w$
and $m_\nu$. The errors in Table II are what would result from a simultaneous
fit to these parameters plus $\Omega_b h^2$, $\Omega_m h^2$, 
$\theta_s$ (the angular
size of the sound horizon), the Helium mass fraction and $\tau$.  
We assume the mean spatial curvature
is zero.  See \cite{kaplinghat03b} for more details.

\begin{table}
\label{tab:cmbexpt}
\begin{center}
\begin{tabular}{ccccccc}
Experiment & $l_{\rm max}^T$ & $l_{\rm max}^{E,B}$ & $\nu$ (GHz) & $\theta_b$ & $\Delta_T$ & $\Delta_P$\\
\tableskip\hline\tableskip
Planck      &2000  & 3000 &  100 & 9.2' & 5.5 & $\infty$ \\
              &    &  & 143 & 7.1' & 6  & 11 \\
              &    &  & 217 & 5.0' & 13 & 27 \\
\tableskip\hline
SPTpol ($f_{\rm sky} = 0.1$) & 2000 & 3000 &  217 & 0.9' & 12 & 17 \\
\tableskip\hline
CMBpol        & 2000 &  3000 & 217 & 3.0' & 1  & 1.4 \\
\tableskip\hline
\end{tabular}
\end{center}
\caption{Experimental specifications.}
\end{table}

\begin{table}
\label{tab:cmb}
\begin{center}
\begin{tabular}{c|c|c|c|c|c}
\tableskip\hline\hline \tableskip Experiment & $m_\nu$ (eV) & $w_x$ &
$\ln P_\Phi^i$ & $n_S$ & $n_S'$ \\
\tableskip\hline\tableskip
Planck  &  0.14&  0.28&  0.016&  0.0074&  0.0032 \\
SPTpol  &  0.11&  0.34&  0.018&  0.01&  0.0057\\
SPTpol + Planck  &  0.082&  0.22&  0.016&  0.0057&  0.0027\\
CMBpol &  0.031&  0.088&  0.011&  0.0024&  0.0014\\
\tableskip\hline
\end{tabular}
\end{center}
\caption{Standard deviations expected from Planck, SPTpol and CMBpol.
From \cite{kaplinghat03b}.}
\end{table}

Are these highly precise measurements of $n_S$ and $n_S'$ valuable?
The difference $n_S-1$ is first order in the slow-roll parameters and
$n_S'$ is second order.  Thus we expect $|n_S'|$ to be on the order
of $(n_S-1)^2$.  If $n_S-1$ = 0.03, which is perfectly consistent with
present data, then we would expect to see non-zero $n_S'$ at about the
$10^{-3}$ level.  It would be tremendously exciting to actually confirm 
this expectation.  Note though that there are models with $|n_S'| \sim n_S-1$
\cite{dodelson02} so this is not a firm test.  
Nevertheless, finding $\sigma(n_S') >> n_S-1$ would be difficult
to reconcile with inflation \cite{gold03}.

\section{CMB+Cosmic Shear}

Adding cosmic shear to CMB data can improve constraints of cosmological
parameters, as was studied in the first CMB+cosmic shear forecasting
paper \cite{hu99a}.  A number of other studies have followed, extending
to use of photometric redshifts to allow tomography \cite{hu99b} and
to examine constraints possible on $w_x$ in addition to the neutrino
mass \cite{hu02d,abazajian02,heavens03}.  Here we focus on errors
on $n_S$ and $n_S'$, though also make some comments about constraints
possible on $m_\nu$ and $w_x$.

The shapes of galaxies are distorted by gravitational lensing.  If we
knew the shapes of the galaxies in the absence of lensing we could
then infer the lensing convergence, $\kappa$,  from the lens--induced 
alteration of
the shape.  Although we have no way of knowing the unlensed shape of
an individual galaxy, we expect that an average over a large number of
galaxy images would be perfectly cylindrical.  The average departure
from a completely cylindrical shape can thus be used as a measure of
the amount of lensing.  From proper analysis of the average galaxy
shape in some pixel, the lensing convergence, $\kappa$, can be recovered.

We model the data ${\bf \kappa}$ (the convergence vector with
index running over $l$, $m$ and redshift bin $z$) as a Gaussian
random field with zero mean and contributions to the total
variance from both signal and noise:
\be
\langle {\bf \kappa} {\bf \kappa}^\dagger \rangle = {\bf S} + {\bf N}.
\ee 
The signal covariance has the structure
\be
S_{lmz,l'm'z'} = C_l^{zz'} \delta_{ll'}\delta_{mm'}
\ee
and the noise covariance is entirely diagonal:
\be 
N_{lmz,l'm'z'} = N_l^{z}\delta_{ll'}\delta_{mm'}\delta_{zz'}.
\ee 

The average ellipticity of galaxies, $\gamma_{rms}$, leads to an
error in the estimated real-space convergence map with variance 
$\gamma_{\rm rms}^2/N^{\rm pix}_{z}$ where $N^{\rm pix}_z$ is the 
mean number of galaxies
in each pixel in redshift bin $z$.  This error is uncorrelated from pixel
to pixel and from redshift bin to redshift bin.  It leads to
an error in the Fourier--transformed convergence with variance 
\cite{kaiser92}
\be
N_l^z = \gamma^2_{\rm rms}\Omega_{\rm pix}/N_z^{\rm pix}.
\ee

For specificity we take $\Omega_{\rm pix} = 11$ sq. arcmin and
$\gamma_{\rm rms} = 0.2$.
The mean number of galaxies in each pixel we take to be 
\be N^{\rm pix}_{z} = 14.2 z^{1.1}
\exp[-(z/1.2)^{1.2}] \ee for $\Delta z = 0.4$.  This is a fit to the
extrapolation from Nagashima et al. 2002 assuming a limiting magnitude
in R of 26 (Tony Tyson, private communication).  We also assume that
half of these galaxies in the $1.2 < z < 2.5$ range we will not be
able to use because we will not be able to get sufficiently accurate
photometric redshifts.  

As the pixel size increases, the number of galaxies increases
and the statistical error drops (equivalently, in Fourier space, 
$l^2 N_l$ increases with $l$ since $N_l$ is constant).  We do
not expect any observations to keep the systematic errors
smaller than $10^{-4}$ on scales above 10'.  With our modeling
of the statistical errors, the rms statistical error on the
shear drops below $10^{-4}$ on scales of 5 degrees.  For this
reason, we quote our results without $l < l_c$ where $l_c = 180/5 = 36$.

The auto- and cross- angular power spectra for the convergence,
$C_l^{zz'}$, are given by (e.g., \cite{hu02a}) 
\be
C_l^{zz'} = {\pi^2 l\over 2} \int dr r W^z(r)W^{z'}(r)\Delta_\Phi^2(k,r)
\ee
with $l=kr$ and
\be
W^z(r) = {1\over \bar{n}_i} {2 \over r} \int_{r(z_{i1})}^{r(z_{i2})} dr' {(r'-r)\over r'} N_i(z') dz'/dr'
\Theta(r'-r)
\ee
where $z_{i1}=z_i-\Delta z/2$, $z_{i2}=z_i+\Delta z/2$ and ${n}_i$ is
\be
\bar{n}_i=\int_{r(z_{i1})}^{r(z_{i2})} dr' N_i(z') dz'/dr'
\ee
and we have assumed that the galaxies are evenly distributed through
the redshift bin.

In addition to the convergence, $\kappa$, there are also the lensing
potential, $\phi$ and the deflection angle ${\bf d}$.  They are
related in real space by ${\bf d} = {\bf \nabla} \phi$ and 
$\kappa = {\bf \nabla}^2 \phi/2$.  Their power spectra are related by 
\be 
2C_l^{\kappa(z)\kappa(z')}/\pi = l(l+1)C_l^{d(\phi)d(\phi)}/(2\pi) = 
l^2(l+1)^2C_l^{\phi(z)\phi(z')}/(2\pi)
\ee

In Fig. 2 we see the convergence power spectra for four redshift
slices: $z=0.6$, $z=1$, $z=2.2$ and $z=1100$.  Errors on $z=1100$ are
those expected from Planck.  Errors on the others are those expected
from the 30,000 sq. degree LSST survey described above with bins in redshift
of width $\Delta z = 0.4$.  The convergence rises monotonically with
redshift since the lensing contribution from each redshift interval
adds incoherently to that of the previous redshift interval.

\begin{figure}[htbp]
  \begin{center}
    \plotone{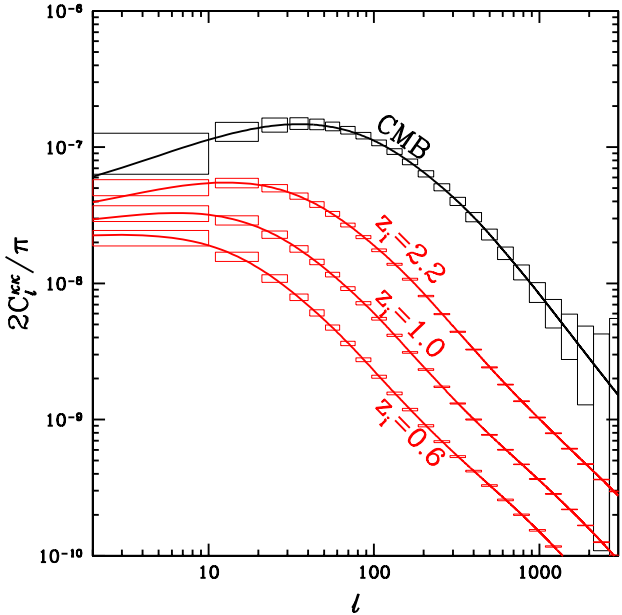}
    \caption{Convergence power spectra for four redshift
slices: $z=0.6$, $z=1$, $z=2.2$ and $z=1100$.  Errors on $z=1100$ are
those expected from Planck.  Errors on the others are those expected
from a 30,000 sq.  degree survey to $m_R = 26$ with bins in redshift
of width $\Delta z = 0.4$.}
\end{center}
\end{figure}

In Fig. 3 we plot the correlation, $r_l^{zz'} \equiv
C_l^{zz'}/\sqrt{C_l^{zz}C_l^{z'z'}}$, for $z=0.6, 1$ and 2.2 with $z'$
fixed to 1100.  The correlation increases with increasing redshift as
the window functions become better matched.  In other words, the
correlation rises with increasing redshift since as $z$ goes towards
1100 a greater fraction of the lensing structures is common
to both.

\begin{figure}[htbp]
  \begin{center}
    \plotone{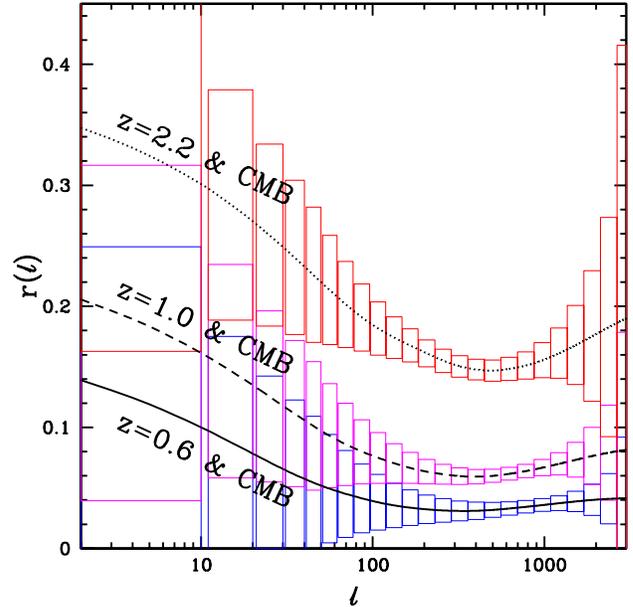}
    \caption{The correlation, $r_l^{zz'} \equiv
C_l^{zz'}/\sqrt{C_l^{zz}C_l^{z'z'}}$, for $z=0.6, 1$ and 2.2 with $z'$
fixed to 1100.  Errors are those expected from Planck plus a 30,000 sq.
degree survey to $m_R=26$ with bins in redshift of width $\Delta z = 0.4$.
}
\end{center}
\end{figure}

By differencing the auto power spectra with appropriate weightings one
can isolate how much lensing power is coming from each redshift range,
and therefore the power spectrum of the matter in each redshift range.
To forecast errors on the parameters that govern the matter power
spectrum, we need not do this differencing.  Our statistical model of
the data is completely specified by the auto and cross--power spectra
and we need only understand our errors on these and how they vary as
we vary the model parameters.

To forecast errors we calculate the Fisher matrix
\bea
F_{pp'} &=& {1\over 2}{\rm Tr}\left[{\bf S}_{,p}{\bf W}{\bf S}_{,p'}{\bf W}\right] \\
& =& \sum_{l,z1,z2,z3,z4} {2l+1 \over 2} C_{l,p}^{z1,z2} W_l^{z2,z3}
C_{l,p'}^{z3,z4} W_l^{z4,z1}
\eea
where the subscript $,p$ means differentiation with respect to parameter
$a_p$ and
\be
{\bf W} \equiv ({\bf S} + {\bf N})^{-1}.
\ee
The inversion of ${\bf W}$ is numerically tractable because
of its block--diagonal structure:
\be
W_{lmz,l'm'z'} = W_l^{zz'} \delta_{mm'}\delta_{zz'}.
\ee
 
Since photometric redshifts are completely unproven for $z > 3$ we
make no use of galaxies at $z > 3$.  To reduce sensitivity to the
non-linear evolution which we may not have modeled well enough we
restrict ourselves to $l < 1000$.  Results are shown in 
Table III.

\begin{table}
\label{tab:cmblsst}
\begin{center}
\begin{tabular}{c|c|c|c|c|c}
\tableskip\hline\hline \tableskip Experiment & $m_\nu$ (eV) & $w_x$ &
$\ln P_\Phi^i$ & $n_S$ & $n_S'$ \\
\tableskip\hline\tableskip
WMAP  &  1.2 &  2.4 &  0.061 &  0.055&  0.018\\
Planck (no lensing) &  0.32&  0.45&  0.016 &  0.0077 &  0.0035 \\
WMAP+LSST  &  0.085 &  0.023 &  0.023 &  0.0094 &  0.0027\\
Planck+LSST &  0.041 &  0.019 &  0.012 &  0.0042&  0.0021\\
\tableskip\hline
\end{tabular}
\end{center}
\caption{Standard deviations expected from WMAP (after 2 years), 
Planck (without reconsruction of $\cldd$), WMAP + LSST and 
Planck (with reconstruction of $\cldd$) + LSST.}
\end{table}

There is not much to be gained by going deeper, at least with our
$z$-independent $l$ cutoff at $l=1000$.  The reason is that the
number density of galaxies is high enough in each redshift bin that
the $C_l$ measurements are sample-variance limited for $l < 1000$.
For the lower redshift bins there is not much to be gained by going to
higher $l$ because that puts us into the fully non-linear regime which
has little to no shape information.  Also in the non-linear regime our
formalism for forecasting the uncertainty breaks down.  But for higher
redshift bins one can in principle push to higher $l$ and gain by it.
The $l$ that divides linear from non-linear is higher at high redshift
for two reasons: 1)a given $k$ projects into a higher $l$, and 2) the
linear to non-linear transition is at higher $k$.  In principle our
$l$ cutoff should be $z$-dependent but we use an $l$-independent one
for simplicity.

There is also not much to be gained by going to higher redshift.
There are two reasons for this.  First, the angular
diameter distance does not change very much over this range of redshift
so we are not probing significantly different spatial scales.  And second,
we are already close to the sample variance limit so the noise reduction due to
the extra galaxies is not useful.  Again, this conclusion could change
if we allowed a $z$-dependent $l$ cutoff that followed the linear/non-linear
transition.  

In addition to the precision determination of the primordial
perturbation spectrum, we can also measure the dark
energy equation-of-state parameter $w_x$ and the neutrino mass to
extraordinary precision.  With data that can measure $w_x$ to 0.02 we
could clearly also place strong constraints on its evolution.  Note, however,
that $m_\nu$ is no better determined than it can be by the CMB alone 
(see Table II) with CMBpol.    

In the first cosmic shear + CMB forecasting papers, \cite{hu99a} and
\cite{hu99b}, it was shown that combining Planck and all-sky cosmic shear
could achieve neutrino mass errors of 0.04 eV without tomography and
0.02 eV with tomography, with $w_x$ fixed to $-1$.  We find similar
results when we fix $w_x$ to -1.  More recently \cite{hu02d},
\cite{abazajian02} and \cite{heavens03} also studied constraints on
neutrino mass and/or $w_x$ that can
come from tomographic cosmic shear surveys.  These studies are done
with all other cosmological parameters held fixed.  The justification
is that these parameters will be determined by CMB observations.  We
find, however, that the remaining uncertainties in cosmological
parameters, even given the expected CMB data, are significant.  This
remaining uncertainty greatly increases errors on $m_\nu$ in
particular, and to a lesser degree on $w_x$.

It may not be possible to control systematic errors at the requisite
levels on large angular scales.  We coarsely simulate the effect of
systematic errors on large angular scales by imposing a low $l$ cutoff
at $l=36$; i.e., we throw out all cosmic shear information at $l < 36$
(though we keep the CMB--determined convergence map all the way down
to $l=2$).  The resulting error increase from this cut is small.  Most
of the weight is coming from higher $l$.  Although we need full sky to
get these exquisitely small errors, we only need full sky to beat down
sample variance, not to accurately determine flucutation power on
large angular scales.

\section{CMB + High-volume redshift surveys}

The errors on $n_S'$ from WMAP+LSST or from higher--resolution
CMB observations, are just at the level of $(n_S-1)^2$ if $n_S \simeq 0.95$.
These experiments may very well leave us only with tantalizing one or two
$\sigma$ determinations of $n_S'$.  We may want to do better.

Better CMB experiments will not do it.  Better tomographic cosmic shear
experiments probably will not do it.  We can understand why with a simple
expression for the error on a power-spectral index, $n$, given errors
on the power $P_1$ and $P_2$ at two different scales, $k_1$ and $k_2$:
\be
\sigma^2(n) = {\sigma^2 (\ln{P_1})+  \sigma^2 (\ln{P_2}) \over [\ln(k_2/k_1)]^2}.
\ee
First note that gaining dynamic range (increasing $k_2/k_1$) only
reduces the error on $n$ logarithmically.  Extending dynamic range is
not a way to dramatically improve the constraints on $n$.  The only
way to dramatically reduce $\sigma(n)$ is to make better measurements
of the power over around a decade (or more) in wavenumber.
2-dimensional surveys can not do this.  Even with 9 uncorrelated
measures of fluctuation power in redshift bins between zero and
last-scattering we can reduce the CMB alone errors by at most a factor
of 3.  

Thus we explore the constraining power of three--dimensional measurements of
clustering.  In particular, we consider a spectroscopic redshift survey.
See \cite{eisenstein03} for more discussion of future redshift surveys.
For simplicity, we ignore effects of redshift distortions and scale-dependent
galaxy bias, which should be included in a more detailed study.   
Note that very deep cosmic shear surveys may be invaluable for calibrating
the galaxy-density relationship for these spectroscopic surveys (in order
to constrain the bias), and for providing the targets for spectroscopic 
follow-up.

For a survey of volume $V$, the error on
the power in a band of width $\Delta k$ is \cite{feldman94,tegmark97d}:
\be
\Delta P(k) = \sqrt{2 \over N_k} \left(P(k)+{1\over \bar n b^2}\right)
\ee
where $N_k$ is the number of independent samples of $\delta({\bf k})$
with $k-\Delta k/2 <|{\bf k}|<k+\Delta k/2$ and is given by
\be
N_k = 4\pi k^2 \Delta k {V \over (2\pi)^3} = k^2\Delta k/(2\pi^2) V.
\ee
The mean number density of galaxies, $\bar n$, determines the
shot-noise contribution to the variance, $1/(\bar nb^2)$ where
$b$ is the galaxy bias such that $P_g(k) = b^2 P(k)$.  

An optimal survey with fixed number of galaxies $N_g = \bar n V$ 
(optimal from a solely statistical error point of view)
will have a volume such that $1/\bar n = b^2 P(k)$ 
on the scale of interest where 
$b$ is the bias of the galaxies.  In this case
\be
{\Delta P(k)\over P(k)} = \sqrt{(8/N_k)} = 2\sqrt{2}\left[N_g b^2 \Delta^2(k) 
\left({\Delta k \over k}\right)\right]^{-1/2}
\ee
where $N_g = \bar n V$ is the number of galaxies in the survey 
and $\Delta^2(k) \equiv k^3 P(k)/(2\pi^2)$.  

Solving for $N_g$ we find 
\be
N_g = 1.5\times 10^9 \left[{10^{-4}\over \Delta P/P}\right]^2\left[{0.5 \over b^2\Delta^2(k)}\right]\left[{k\over \Delta k}\right].
\ee
This is a lot of galaxies.  And unfortunately, most of them are at high
redshift (if we want $\Delta P/P = 10^{-4}$).  If the volume is
spherical with coordinate distance radius of $R$ then
\be
R = 7 h^{-1} Gpc {\left(0.2 h Mpc^{-1} \over k \right)} (k/\Delta k)^{1/3} \left[{10^{-4}\over \Delta P/P}\right]^{2/3}.
\ee
For reference, the distance to the horizon is about 14 Gpc \cite{knox01,spergel03}.

One might conceivably do better by extending to higher $k$ than the
fiducial 0.2 $h$ Mpc$^{-1}$.  The main benefit is the reduced volume
one has to sample ($V \propto k^{-3}$, see above equation) to
achieve the same value of $\Delta P/P$.  The higher this is pushed,
however, the more one must worry about scale-dependent bias.  And
when one wants $\Delta P/P$ as small as $10^{-4}$ even a tiny
amount of unknown scale-dependence can lead to highly significant
systematic error.

\section{Conclusions}

If inflation happened at sufficiently high energy scales 
($V^{1/4} \ga 2 \times 10^{15}$ GeV) and nature is kind to
us with respect to astrophysical foregrounds (dust, synchroton
radiation, etc.) then we can detect the influence of tensor
perturbations in the CMB polarization.  If the energy scale is
slightly higher, we will be able to verify (or rule out) the
inflation consistency equation.

For measurement of the scalar perturbation spectrum, constraining
$n_S$ and $n_S'$ to better than $10^{-3}$ may be impossible.  This
level can be reached with a post-Planck CMB polarization mission
or by combining WMAP or Planck with all-sky tomographic cosmic shear
observations.  Spectroscopic redshift surveys, even with benign
assumptions about galaxy bias, are unlikely to improve our ultimate
constraints on $n_S$ and $n_S'$.  

\section{Acknowledgments}
First and foremost I thank my collaborators, Yong-Seon Song and 
Manoj Kaplinghat,
who did almost all the work presented here.  I also thank A. Albrecht, B.
Gold, N. Kaloper, D. Spergel, A. Tyson and D. Wittman for useful conversations.

\bibliography{cmb3}

\end{document}